\documentclass[journal=apchd5,manuscript=letter,layout=twocolumn]{achemso}
\setkeys{acs}{keywords = true}
\setkeys{acs}{articletitle = true}
\begin{tocentry}
 \includegraphics[scale=0.75]{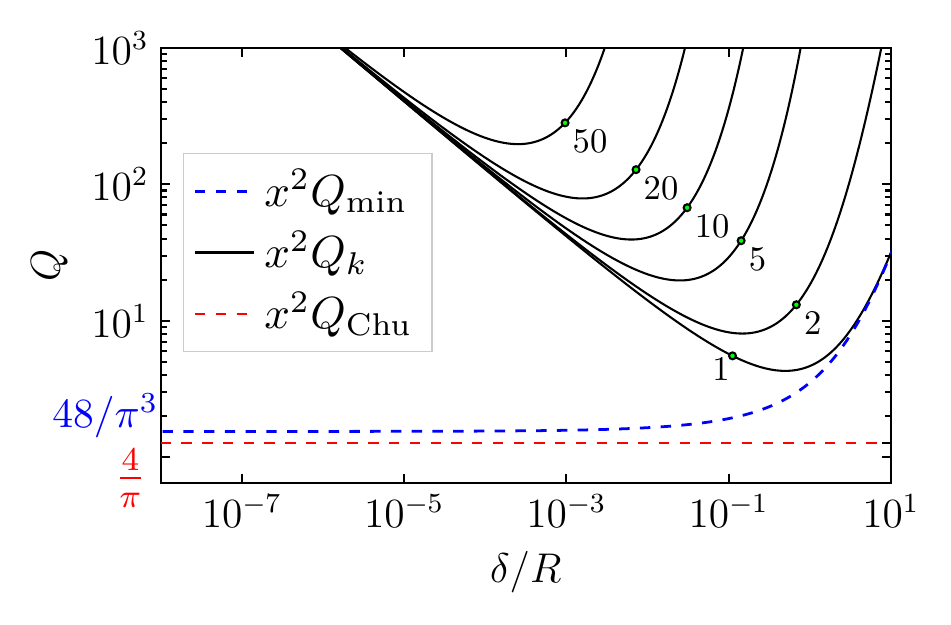}
\end{tocentry}
\usepackage[version=3]{mhchem} 
\usepackage{color}
\usepackage{multirow}
\usepackage{lipsum}
\usepackage{color, soul}
\usepackage{esint}

\usepackage[utf8]{inputenc}
\usepackage{setspace,amsmath,amsfonts}
\usepackage{graphicx}
\usepackage{systeme}
\usepackage{epstopdf}
\usepackage{tikz,pgfplots}
\usetikzlibrary{shapes,decorations}
\usepgfplotslibrary{fillbetween}
\pgfplotsset{compat=newest}
\usepgfplotslibrary{groupplots}
\usepgfplotslibrary{dateplot}
\usepackage [autostyle, english = american]{csquotes}
\MakeOuterQuote{"}
\usepackage{hyperref}

\usepackage{mathrsfs}
\usepackage{eso-pic}
\usepackage{amssymb}

\DeclareMathOperator{\arccosh}{Arccosh}

\newcommand{\llangle}[1][]{\savebox{\@brx}{\(\m@th{#1\langle}\)}%
  \mathopen{\copy\@brx\kern-0.5\wd\@brx\usebox{\@brx}}}
\newcommand{\rrangle}[1][]{\savebox{\@brx}{\(\m@th{#1\rangle}\)}%
  \mathclose{\copy\@brx\kern-0.5\wd\@brx\usebox{\@brx}}}
\makeatother

\newcommand{\secTD}{\Omega}

\newcommand{\tOmega}{ \tilde{\Omega} }

\newcommand{\re}[1]{\text{Re}\left\{ #1 \right\}}
\newcommand{\im}[1]{\text{Im}\left\{ #1 \right\}}

\newcommand{\rb}{\mathbf{r}}

\newcommand{\rrp}{\left( {\mathbf{r}} \right)}
\newcommand{\rrpp}{\left( {\mathbf{r}}' \right)}

\author{Mariano Pascale}
\affiliation{ Department of Electrical Engineering and Information Technology, Universit\`{a} degli Studi di Napoli Federico II, via Claudio 21,
 Napoli, 80125, Italy}
 \alsoaffiliation{Photonics Initiative, Advanced Science Research Center, City University of New York, New York, New York 10031, USA}
\author{Sander A. Mann}
\affiliation{Photonics Initiative, Advanced Science Research Center, City University of New York, New York, New York 10031, USA}
\author{Carlo Forestiere}
\affiliation{ Department of Electrical Engineering and Information Technology, Universit\`{a} degli Studi di Napoli Federico II, via Claudio 21,
 Napoli, 80125, Italy}
 \author{Andrea Alù}
 \affiliation{Photonics Initiative, Advanced Science Research Center, City University of New York, New York, New York 10031, USA}
 \alsoaffiliation{Physics Program, Graduate Center of the City University of New York, New York, New York 10016, USA}
 \email{aalu@gc.cuny.edu}
 \title{On the bandwidth of singular plasmonic resonators in relation to the Chu limit}

\begin{document}

\onecolumn
\begin{abstract}
Plasmonic nanostructures with singular geometries can exhibit a broadband scattering response that at first glance appears to violate the lower bounds for the radiation quality ($Q$) factor of small radiators, known as the Chu limit. Here we explore this apparent contradiction, investigating  the $Q$ factor of the resonant modes supported by two nearly touching  cylinders, and analyze how their fractional bandwidth fares in relation to the Chu limit. We first derive lower bounds for the radiation $Q$ factors of two-dimensional objects of arbitrary cross-section. We then discuss the dissipation and radiation $Q$ factors associated with the plasmonic resonances of a cylinder dimer as a function of its gap size. We show that the radiation $Q$ factor is always larger than the minimum $Q$ and, as long as the peaks in the scattering spectrum are well separated, their bandwidth is equal to the inverse of their $Q$ factor. In the limit of touching cylinders, the resonance spectra transition from discrete to a continuum around an accumulation point, yielding a broadband response for any finite level of material loss. Within any given frequency interval, the response is the result of a multitude of plasmon resonances, each individually obeying the Chu limit. Nevertheless, the connection between the $Q$ factor and the overall bandwidth of the scattering response is lost. Our study sheds light onto the exotic resonant phenomena emerging when plasmonic materials are shaped in singular geometries, and outlines their opportunities and limitations for nanophotonics.

\end{abstract}
\twocolumn

\newif\iffig
\figtrue
Achieving strong light-matter interactions within a small volume is a prominent goal of the field of photonics \cite{Koenderink2015}. Impressive results have been demonstrated using plasmonic \cite{fischer_engineering_2008,Kuttge2010,Ciraci2012} or high-index dielectric resonators \cite{Schuller2009a,Huang2013,groep_direct_2016,Kapitanova2017,Yang2017,rybin_high-q_2017}, with applications in, for example, nonlinear optics \cite{Lee2014,koshelev_subwavelength_2020}, photovoltaics \cite{Brongersma2014,Mann2016}, photoluminescence \cite{Kinkhabwala2009,Akselrod2014}, and sensing \cite{Bosio2019}. However, shrinking the footprint of a resonant device generally comes at the cost of bandwidth (and therefore, e.g., operational speed). This fundamental trade-off between volume, peak field enhancement or scattering, and bandwidth has been investigated using various analytical and numerical methods \cite{gustafsson_physical_2007,Miller2013a,kuang_2020,thal_new_2006,gustafsson_physical_2015,vandenbosch_simple_2011}, the best known of which is probably the Chu limit on the bandwidth of small radiators \cite{chu_physical_1948,collin_evaluation_1964,mclean_re-examination_1996}. 

The Chu limit provides a lower bound for the minimum radiation quality ($Q$) factor of a small antenna. This limit applies to both self-resonant small objects, including plasmonic and high-index nano antennas, and externally tuned objects such as small radio-frequency antennas. For three dimensional (3D) electric radiators, e.g., subwavelength plasmonic resonators, the minimum radiation $Q$ factor is \cite{thal_new_2006,vandenbosch_simple_2011,gustafsson_physical_2015},
\begin{equation}
    Q_\text{min} = \frac{1.5}{\left(k a\right)^3},
    \label{eq:ChuEle}
\end{equation}
where $a$ is the radius of the minimum sphere circumscribing the object, $k=\omega/c_0$ is the wavevector in vacuum,  $\omega$ is the resonance frequency, and $c_0$ is the speed of light in vacuum. Similarly, for magnetic radiators, e.g., subwavelength dielectric resonators, the minimum radiation $Q$ factor is \cite{gustafsson_physical_2015,thal_new_2006,vandenbosch_reactive_2010}
\begin{equation}
        Q_\text{min} = \frac{3}{\left(k a\right)^3}.
    \label{eq:ChuMag}
\end{equation}
Recently, stricter bounds that depend on the shape of the enclosing volume have been also introduced \cite{thal_new_2006,gustafsson_physical_2015,vandenbosch_simple_2011}. Analogous lower bounds for the $Q$ factor of two-dimensional (2D, translationally invariant along one dimension) radiators have not been derived yet, despite some preliminary work in this context \cite{collin_evaluation_1964}.

While according to Eq. (1) a resonance contained within a sphere with normalized radius $ka$ must have a $Q$ factor exceeding $1.5/(ka)^3$, deeply subwavelength, yet extremely broadband plasmonic structures have been recently reported \cite{aubry_plasmonic_2010,fernandez-dominguez_collection_2010,Lei2010,Pendry2012}. These objects are typically characterized by singular geometries, such as touching spheres \cite{schnitzer_asymptotic_2020} or cylinders \cite{bladel_2010}, and appear to be seemingly at odds with the Chu limit. 

Here, we reconcile these contrasting results, and discuss in detail the origin of the broad bandwidth in these structures, focusing on a touching dimer of cylinders as a model system \cite{aubry_plasmonic_2010,fernandez-dominguez_collection_2010}. To place it in the context of the Chu limit, we first derive the analogue of the Chu limit for 2D systems in Sec.~\ref{sec:Qbounds}. Then, in Sec. \ref{sec:Resonances} we present analytical formulas for the dissipation and radiation $Q$ factors of the bright modes of such a dimer, providing quantitative criteria to identify the dominant damping mechanisms. In Sec.~\ref{sec:Link}, we investigate  the  absorption  power  spectrum of two nearly touching cylinders excited by a plane wave, highlighting the phenomena underlying the broadband resonance observed in the absorption spectrum when the dimer gap vanishes. Finally, in Sec.~\ref{sec:Disorder}, we discuss how the extreme bandwidth of such singular structures makes them highly susceptible to disorder and imperfections.

\section{The minimum $Q$  factor of 2D radiators}
\refstepcounter{section}
\label{sec:Qbounds}
 In this section, we derive the minimum $Q$ factor that can be achieved by a current density ${\bf j}$ supported by 2D radiators of small cross-section $\Omega$. Existing literature provides $Q$ factor bounds for radiators belonging to two disjoint categories: radiators of the electric kind and radiators of the magnetic kind, characterized by their induced current densities. Radiators of the electric type support currents with zero curl, i.e., {\it longitudinal} currents, while radiators of magnetic type support currents with zero divergence, i.e., {\it transverse} currents. In the case of subwavelength particles, it is useful to classify these radiators by their platform: small plasmonic particles support longitudinal currents, while small dielectric particles support transverse currents \cite{forestiere_resonance_2020}. In the following, we will follow this distinction.
 
\subsection{Radiator of electric kind}
\begin{figure}[tpb!]
\centering
\includegraphics[width=0.99\columnwidth]{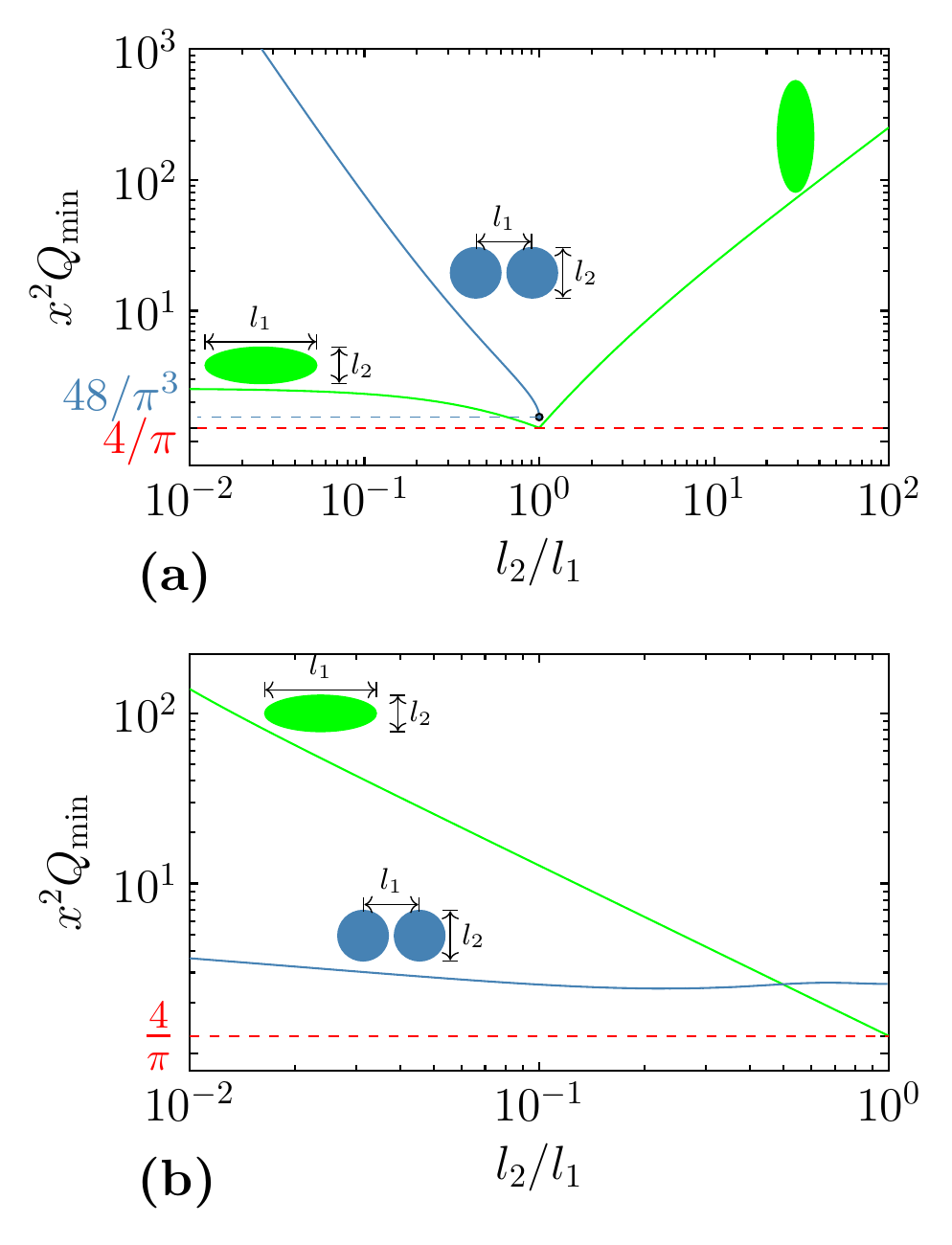}
\caption{Lower bounds for $ x^2 Q$ for 2D radiators of electric  $\bf (a)$ and magnetic type $\bf (b)$, as a function of geometrical parameters (see inset). They are a cylinder with ellipsoidal cross-section (green line) and a pair of circular cylinders  (light blue line). For the ellipsoid $l_1$ and $l_2$ are the length of the axes, for the pair of cylinders  $l_1$ is the center-to-center distance, $l_2$ is the diameter. 
}
\label{fig:Qsumm}
\end{figure}
\noindent
For small radiators, we may consider the electromagnetic problem as quasistatic. For a radiator of electric kind, the $Q$ factor is expressed as $2 \pi$ times the ratio of the electrostatic energy stored in the electromagnetic field to the energy radiated to infinity in a single period,

\begin{equation}
        \mathcal{Q}  = - \frac{8}{\pi}  \frac{\displaystyle  \oint_{ \partial \secTD}  \sigma  \rrpp \oint_{ \partial \secTD} { \sigma  \rrp } \log \Delta r \, dl dl'}{\displaystyle \oint_{\partial \secTD} \sigma   \rrpp  \oint_{\partial \secTD} \sigma \rrp   (\Delta r)^2 \, dl dl'} \, \frac{1}{x^2},
        \label{eq:Qel}
\end{equation}
where $\partial \secTD$ is the object cross-section boundary, $\hat{\bf n}$ is the outward-pointing normal unit vector lying in the cross-sectional plane, $\sigma = {\bf j}\cdot \hat{\bf n}$ is the surface charge density per unit length (p.u.l.), and $\Delta r = \left|\rb - \rb '\right|/\ell_c$, with $\ell_c$ being a characteristic linear length of the radiator cross-section. The size parameter $x$ is defined as $\displaystyle x=\frac{\omega}{c_0}\ell_c$. From now on, we shall assume $\ell_c$ to be the radius of the smallest circle enclosing the radiator cross-section.

To derive the Chu limit for this scenario, we need to minimize Eq.~\ref{eq:Qel}. Following \cite{gustafsson_physical_2015}, the minimization can be recast as finding the {\it optimal} current distribution with zero curl, with a specified, i.e., constrained, squared magnitude of the electric dipole moment, which yields the minimum electrostatic energy stored in the whole space. 
The minimum of the $Q$ factor is then obtained as \cite{gustafsson_physical_2015}

\begin{equation}
    \left( x^2 Q \right)_\text{min} =  \frac{8}{\gamma_\text{max,e}},
    \label{eq:QminEQS}
\end{equation}
where $\gamma_\text{max,e}$ is the largest of the 2 eigenvalues of the electric polarizability tensor $\boldsymbol{\gamma}_e$ of the radiator, scaled by $l_c^2$. The electric polarizability tensor of a 2D object is defined in the Methods.

Let us apply this scheme to an infinite cylinder of circular cross section. Due to symmetry, its eigenvalues are degenerate and given in the Methods.
The resulting minimum $Q$ factor is
\begin{equation}\label{eq:minQEQS}
Q_\text{min} =\frac{4}{\pi}\frac{1}{x^2},
\end{equation}
and it represents the equivalent of the Chu limit for 2D objects. Using this approach, we can also calculate the minimum $Q$ factor of a current distribution constrained within an elliptical cross-section, as a function of the eccentricity $l_2/l_1$, with $l_1,\,l_2$ the elliptical cross-section axes (see the inset in Fig. \ref{fig:Qsumm}(a)). In this case, the cross section has reflection symmetry, and the electric polarizability tensor $\boldsymbol{\gamma}_e$ can therefore be cast as a diagonal matrix by choosing an appropriate coordinate system. The size parameter $x$ is $\displaystyle x=\frac{\omega}{c_0}\frac{\max{\{l_1,l_2\}}}{2}$.
We compare the resulting minimum $Q$ for the ellipsoidal cross section (in green) in Fig.  \ref{fig:Qsumm}(a) with the one of a cylinder (in red, dashed). The cylinder is polarized along a fixed direction, resulting in the asymmetry around $l_1=l_2$.

Fig. \ref{fig:Qsumm}(a) also shows the minimum $Q$ factor of a current distribution constrained within two coupled infinite cylinders of circular cross section, as a function of the ratio $l_2/l_1$, where $l_2$ is the diameter of one of the cylinders, and $l_1$ is the center-to-center distance.
Here, the size parameter $x$ is $\displaystyle x=\frac{\omega}{c_0}\frac{(l_1+l_2)}{2}$.
In this case, when $l_1$ approaches  $l_2$, the edge-edge gap between the two cylinders becomes very small and $\left(x^2Q\right)_\text{min}$ tends to a finite value $\left(x^2Q\right)_\text{min} \rightarrow 48/\pi^3$. Hence, the $Q$ factor of any longitudinal current density distribution supported by a pair of cylinders has to be greater not only than $\displaystyle\frac{4}{\pi}\frac{1}{x^2}$, but also than $\displaystyle\frac{48}{\pi^3}\frac{1}{x^2}$, which is the limiting value for the two cylinders when they touch.

\subsection{Radiator of the magnetic kind}
While in the remainder of this paper we will only deal with 2D plasmonic structures, which belong to the electric type, for the sake of completeness we derive here the minimum $Q$ factor of a current density distribution supported by a radiator of magnetic type as well. In this case, the $Q$ factor may be expressed as $2 \pi$ times the ratio of the magnetostatic energy stored in all of space to the energy radiated to infinity in a period
\begin{equation}
\label{eq:QminMagnFunc2D}
        \mathcal{Q}  =  \frac{8}{\pi}  \frac{\displaystyle  \int_{  \secTD}  {\bf j}  \rrpp \int_{  \secTD} {{\bf j}  \rrp } \log \Delta r \, dS dS'}{\displaystyle \int_{ \secTD} {\bf j}   \rrpp  \int_{ \secTD} {\bf j} \rrp   (\Delta r)^2 \, dS dS'} \, \frac{1}{x^2}.
\end{equation}
Following the same approach as for the electric type, but using the squared magnitude of the magnetic dipole moment, we find that the minimum $Q$ factor is

\begin{equation}
    \left( x^2 Q \right)_\text{min} =  \frac{4}{\gamma_\text{m}},
    \label{eq:QminEQS2D}
\end{equation}
where $\gamma_\text{m}$ is the scalar  magnetic polarizability. The magnetic polarizability of 2D objects is given in the Methods. 
While Eq.~\ref{eq:QminEQS2D} differs from Eq.~\ref{eq:QminEQS}, the result for the minimum $Q$ factor is identical to the electric type:
\begin{equation}
Q_{min} =\frac{4}{\pi}\frac{1}{x^2}.
\end{equation}
Fig. \ref{fig:Qsumm}(b) shows the minimum $Q$ factor of a current distribution confined within a 2D object of elliptical cross-section, as a function of the eccentricity, as well as the minimum $Q$ factor of a current distribution constrained within two coupled cylinders of circular cross section, as a function of the the ratio $l_2/l_1$. Because for radiators of the magnetic kind, the current is a loop on the boundary, the minimum $Q$ factor is now symmetric about $l_1=l_2$.

\section{Bandwidth \& $Q$ factor \\ in singular plasmonic structures}
\refstepcounter{section}
\label{sec:Resonances}
 In this section, we analytically derive the radiation and dissipation $Q$ factors of the resonant modes of a cylinder dimer. The two cylinders, whose cross-section is shown in Fig. \ref{fig:CrossSection}, occupy a domain $\Omega$, are separated by a gap size $\delta$, and have total linear dimension $D$, which is the diameter of the minimum circumscribing circle. The cylinders are made of a linear, homogeneous, isotropic, nonmagnetic, and time-dispersive material with relative dielectric permittivity $\varepsilon_R $. 
 \iffig
\begin{figure}[t!]\label{eq:EsMIM}
 \centering
\includegraphics[width=0.6\columnwidth]{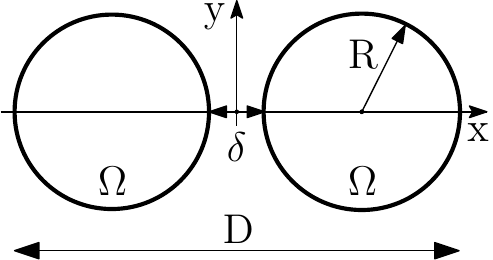} 
\caption{Cross-section of two infinite identical circular cylinders.}
\label{fig:CrossSection}
\end{figure}
\fi

To find the electrostatic (plasmon) resonances of the structure, the problem can be formulated as an eigenvalue problem for a specific linear integral equation, where the spectral parameter is the relative dielectric permittivity \cite{mayergoyz_analysis_2007}. By solving this eigenvalue problem, we determine the structure's resonant relative dielectric permittivities, i.e., {\it eigenpermittivities}, and the electric fields of the corresponding plasmonic modes. The eigenpermittivities and the plasmonic modes are countably infinite.

We assume a time-varying incident electric field, linearly polarized along the dimer axis, and spatially uniform:
\begin{equation}\label{eq:Einc}
 {\bf e}_{inc} \left(t\right)= E_0 f\left(t\right) \, {\bf \hat x},
\end{equation}
where $E_0$ is a real amplitude.  We consider a harmonic excitation, i.e., $f \left( t \right) = e^{ - i \omega t }$. The scattered electric field $\mathbf{e}  \left( t \right) = \text{Re} \left\{ \mathbf{E} \, e^{ - i \omega t } \right\}$ is everywhere defined as the difference between the total $\mathbf{e}_{tot}  \left( t \right) = \text{Re} \left\{ \mathbf{E}_{tot} \, e^{ - i \omega t } \right\}$ and the incident field as $\mathbf{e} = \mathbf{e}_{tot} - {\bf e}_{inc}$. 
Within the quasi-electrostatic approximation, the scattered electric field can be written as \cite{mayergoyz_electrostatic_2005,forestiere_material-independent_2016,pascale_full-wave_2019}

\begin{equation}\label{eq:Efield}
\mathbf{E} \left( \mathbf{r}, \omega \right) = E_0 \left[\varepsilon_R(\omega)-1\right]\sum_{k=1}^{\infty}  \frac{\left\langle \hat{\bf x},{\bf E}_k\right\rangle}{\varepsilon_k  - \varepsilon_R(\omega)}   {\bf E}_k\left(\mathbf{r}\right),
\end{equation}
where $\left\{ \mathbf{E}_k \right\}_{k\in\mathrm{N}}$ are the normalized electrostatic modes, whose expression is given in  the Methods, 
and $\varepsilon_k$ are the corresponding eigenpermittivities, which are in turn the union of two twin sets \cite{mayergoyz_electrostatic_2005,mayergoyz_plasmon_2013,klimov_nano_2015}
\begin{align}
\varepsilon^+_k &= - \coth{k  \mu },\nonumber\\
 \varepsilon^-_k &= - \tanh{k  \mu},
\end{align}
where $ \varepsilon^-_k = 1/\varepsilon^+_k$,
associated to bright  ${\bf E}_k^+$ and dark ${\bf E}_k^-$ modes, respectively, as they exhibit non-zero and zero dipole moments. 
$\left\langle {\bf A},{\bf B} \right\rangle$ is the scalar product $ \int_\Omega {\bf A} \cdot {\bf B} \, dS $, and
\begin{equation}\label{eq:mu}
\displaystyle\mu=\arccosh{\left(1+\frac{\delta}{2R}\right)},
\end{equation}
is the bipolar coordinate of the cylinder dimer boundary, with $\arccosh$ the inverse hyperbolic cosine. Both the eigenpermittivities and the corresponding modes depend on the ratio $\delta/R$, but not on the overall size $D$ or the constituent material, since we are studying a quasi-static scenario. In the limit of very small gaps, i.e., $\delta/R \ll 1$, $\mu$ can be approximated as $\mu \simeq \sqrt{\delta/R}$.
\iffig
 \begin{figure}[tb]
 \centering
\includegraphics[width=0.9\columnwidth]{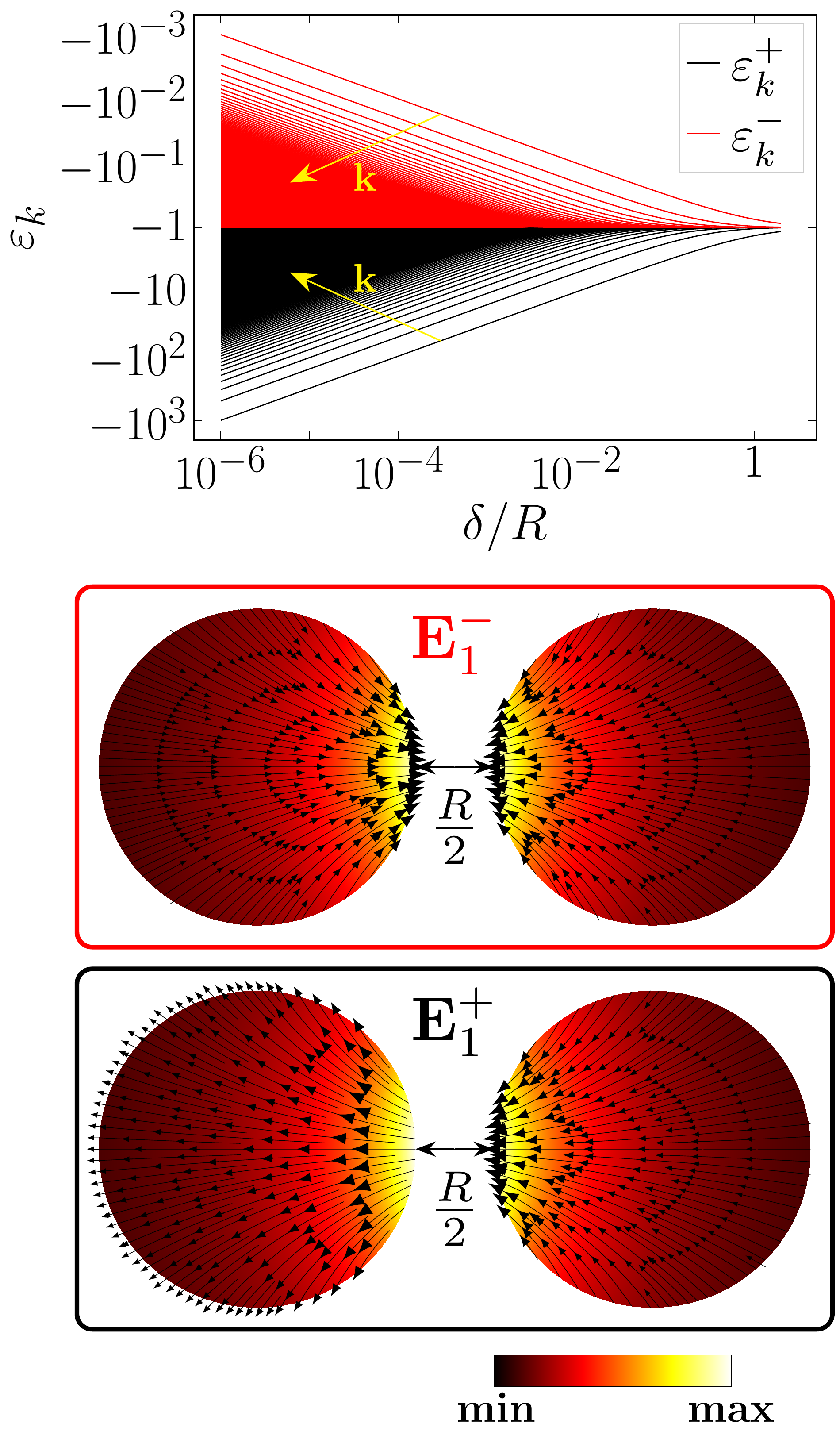}
\caption{Bright ($\varepsilon_k^+$) and dark ($\varepsilon_k^-$) eigenpermittivities as a function of $\delta/R$. The field lines of the electric field of the first bright (${\bf E}_1^+$) and  dark (${\bf E}_1^-$) modes for $\delta/R=\frac{1}{2}$ are shown below.}
\label{fig2}
\end{figure}
\fi

In Fig. \ref{fig2} we plot the eigenpermittivities $\varepsilon^+_k$ (black) and $\varepsilon^-_k$ (red) as a function of the relative gap size $\delta/R$, parametrized by the mode index $k$, and an example of bright and dark modes for $\delta/R=0.5$. For any relative gap size, as $k\rightarrow \infty$ the eigenpermittivities tend to ${\varepsilon_{acc}}=-1$, which is the accumulation point of 2D plasmonic objects regardless of their shape \cite{mayergoyz_analysis_2007}. In the limit of well separated cylinders, i.e., $\delta/R \rightarrow \infty$, all the eigenpermittivities approach $\varepsilon_{acc}$, and the scattering problem reduces to the one of two non-interacting cylinders, whose eigenpermittivities are located at ${\varepsilon_{acc}}$. As the gap size $\delta$ decreases, the bright and dark eigenpermittivities shift toward more negative and positive values, respectively. Lower-order eigenpermittivities shift more than higher-order eigenpermittivities. As we approach the limit $\delta/R \rightarrow 0$, the plasmonic spectrum becomes a continuum \cite{mcphedran_electrostatic_1981},  as the eigenpermittivities of bright modes fill the semiaxis $\left(-\infty, \varepsilon_{acc} \right)$, and the eigenpermittivities of dark modes fill the interval  $\left(\varepsilon_{acc}, 0 \right)$.

To relate the eigenpermittivities to frequencies, we consider a Drude model for the relative dielectric permittivity $\varepsilon_R(\omega)$:
\begin{equation}\label{eq:Drude}
\varepsilon_R = 1- \frac{\omega_p^2}{\omega(\omega+{\rm i}\nu)}.
\end{equation}
We choose $\omega_p=13.07\times 10^{15}\,{\rm rad/s}$ and $\nu=\nu_0=131.19\times 10^{12}\, {\rm rad/s}$, representative of silver \cite{johnson_optical_1972}. The resonance frequencies related to the eigenpermittivities $\varepsilon^+_k$ are then given by

\begin{equation}\label{eq:omegak}
    \omega_k^+ = \frac{\omega_p}{\sqrt{1-\varepsilon^+_k}},
\end{equation}
and the frequency $\omega_{acc}$ corresponding to the accumulation point of the plasmon spectrum ($\mbox{Re}\{\varepsilon_R(\omega_{acc})=-1\}$) is $\omega_{acc} = {\omega_p}/{\sqrt{2}}$. 

The absorption cross section $\sigma_\text{abs}$ is obtained by normalizing the absorbed power per unit length $P_{abs}$, given in Eq. \ref{eq:PabsMIM} of the Methods,
by the incident irradiance $\left( c_0 \varepsilon_0 |E_0|^2/2 \right)$. As a result, $\sigma_\text{abs}$ normalized by the squared circumscribing cylinder diameter $D^2$ is independent of the object size, and with Drude dispersion it has the expression

\begin{multline}
\label{eq2}
\sigma_\text{abs}/D^2  =\left(\frac{\delta/R}{\delta/R+4}\right) \times \\ \times 2\pi \frac{\nu }{c_0}\omega_p^2  \sum_{k=1}^{\infty}   \frac{k\, {\rm e}^{-2 k \mu }}{\omega^2\left[\left(\omega_k^+/\omega\right)^2-1\right]^2 + \nu^2} ,
\end{multline}
where $\mu$ is defined in Eq. \ref{eq:mu}.

The average electric field over the structure section $\Omega$ is equal to the  electric dipole moment ${\bf p}_{k}$ of the mode ${\bf E}_k^+$ \cite{mayergoyz_electrostatic_2005}, i.e., 
\begin{equation}\label{eq:tot_dipole}
 {\bf p}_{k}=\int_\Omega {\bf E}_k^+ \,dS.
\end{equation}
By combining this equation with the expression of the bright plasmonic mode ${\bf E}_k^+$ of the structure in Eq. \ref{eq:modes_expression} of the Methods, 
and normalizing it by the length  $D/2$, we obtain the mode's normalized electric dipole moment ${\bf P}_{k}$
\begin{equation}
{\bf P}_{k} =\frac{1}{D/2} {\bf p}_k=\sqrt{ 8\pi \frac{\delta/R}{\delta/R+4} k}\,{\rm e}^{-k\mu}\,{\bf \hat x},
\label{eq:Dipole2cyl}
\end{equation}
which is independent of the object size.
Here $\mu$ is defined in Eq. \ref{eq:mu}, and ${\bf \hat x}$ is the unit vector directed along the structure axis (see Fig. \ref{fig:CrossSection}). 

In the regime in which the material loss is dominant, the $Q$-factor is given by \cite{wang_general_2006}
\begin{equation}
Q^d =  \frac{\omega}{2\im{\varepsilon_R \left(\omega\right)}}\frac{d\,\,}{d\omega}\re{\varepsilon_R \left( \omega \right) }.
\end{equation}
In particular, for a Drude dispersion relation, as the one in Eq. \ref{eq:Drude}, the $Q$-factor of a resonance at frequency $\omega^+_k$ has the expression
\begin{equation}
\label{eq:QdissDrude}
Q^d_k=\frac{\omega^+_k}{\nu}.
\end{equation}
Since the relative bandwidth of an isolated mode approximates the reciprocal of its $Q$ factor (when large), for a Drude dispersion relation the relative bandwidth is directly proportional to  $\nu$.
In Fig. \ref{fig:QdQr}(a) we plot the dissipation $Q$ factor $Q^d \left( \nu \right)$ as a function of $\nu$ for the two limit cases of low-order ($k=1$) and $\infty$-order plasmonic modes; the remaining curves for finite value $k$ are contained in the region delimited by them. According to Eq. \ref{eq:QdissDrude}, the mode with lowest dissipation $Q$ is the one associated to the lowest resonance frequency. 


\iffig
\begin{figure}[tb]
\centering
\includegraphics[width=\columnwidth]{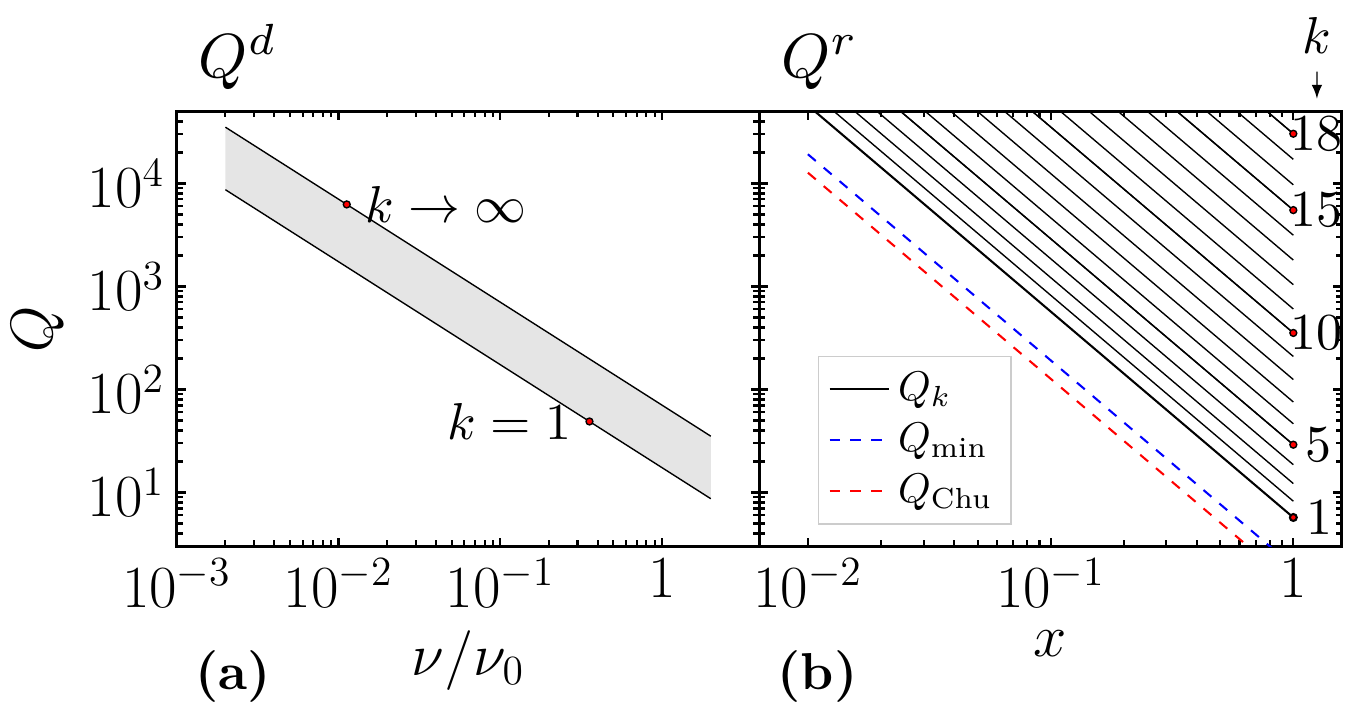}
\caption{Dissipation $Q^d_k$ $\bf (a)$ and radiation $Q^r_k$ $\bf (b)$ factors for a Drude metal cylinder dimer with relative gap size $\delta/R=0.1$ as functions of $\nu/\nu_0$, $\nu_0=131.19\times 10^{12}\, {\rm rad/s}$, and $x=\frac{\omega}{c_0}\frac{D}{2}$, respectively, parametrized by the mode index $k$. The minimum radiation $Q$ achievable by longitudinal currents confined within this structure is shown with a blue dashed line. The minimum radiation $Q$ for all 2D structures, i.e., $Q_\text{Chu} = \frac{4}{\pi}\frac{1}{x^2}$ is also shown with a red dashed line.}
\label{fig:QdQr}
\end{figure}
\fi

The radiation $Q$ factor of the $k^\text{th}$ bright resonant mode has the expression \cite{forestiere_resonance_2020}
\begin{equation}
Q_k^r =\frac{8}{\left|\varepsilon_k-1\right|} \frac{1}{|{\bf P}_{k}|^2} \frac{1}{x^2}=\frac{4+\delta/R}{2\pi k \,\delta/R}\left({\rm e}^{2 k \mu}-1\right) \frac{1}{x^2},
\label{eq:Qfac_Cylinders}
\end{equation}
where we define the size parameter $x$ of the present structure as $\displaystyle x=\frac{\omega}{c_0}\left(\frac{D}{2}\right)$.

In Fig. \ref{fig:QdQr}(b) we plot the radiation $Q$ factor $Q^r_k$ as a function of $x$, for nearly touching cylinders with $\delta/R=0.1$. The curves are parametrized for different values of the mode index $k$. We conclude that, for any given size parameter $x$, higher-order modes correspond to higher values of the radiation $Q^r_k$. In the same panel we also show the minimum $Q$ factor of any two-dimensional radiator of the electric type, denoted as $Q_\text{Chu}$ and the minimum $Q$ of a current distribution confined to a cylinder dimer with $\delta/R=0.1$, denoted as $Q_\text{min}$. As expected, all $Q$ factors lie above both the Chu limit and the more specific dimer limit.

\begin{figure}[tpb]
\centering
\includegraphics[width=0.99\columnwidth]{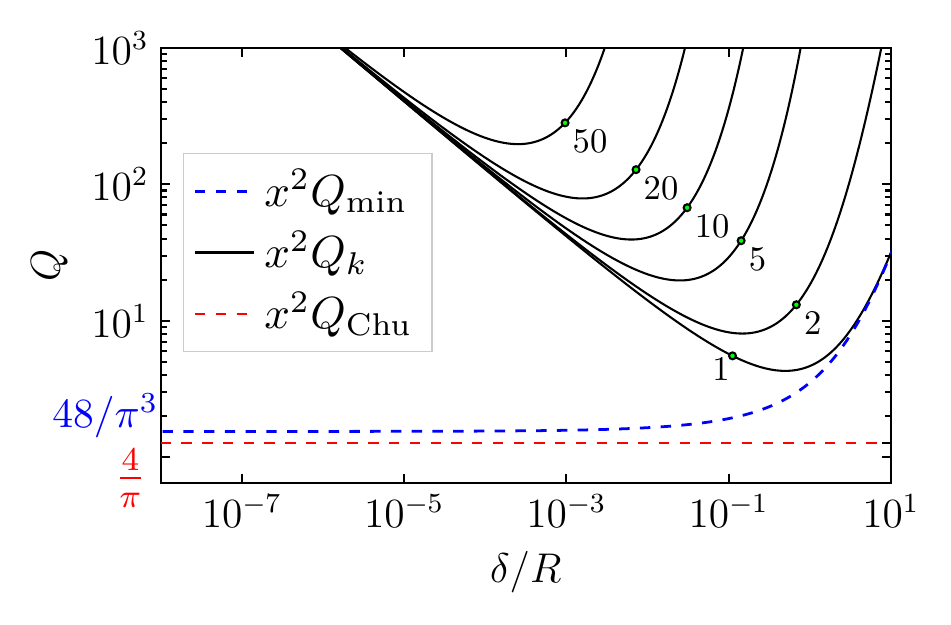}
\caption{Radiation $Q$ factor normalized with $x^2$ of the modes of the cylinder dimer as a function of the gap-to-radius ratio $\delta/R$, parametrized by the modes' index $k$. The minimum $Q$ factor achievable by the structure (blue dashed line) and the Chu limit (red dashed line) are also shown.}
\label{fig:Qk_Qmin}
\end{figure}

In Fig. \ref{fig:Qk_Qmin} we show the radiation $Q$ factor $\left( x^2 Q_k^r \right)$ of the modes with indices $k=(1,2,5,10,20,50)$, as a function of the relative gap size $\delta/R$. They are compared against the minimum $Q$ factor $ \left( x^2 \, Q \right)_\text{min}$ achievable by longitudinal currents constrained within the cylinder dimer (blue dashed line) and against the Chu limit (red dashed line) $\left( x^2 \, Q \right)_\text{Chu}$. 
As the relative gap sizes $\delta/R$ decreases, the $Q$ factor of the individual modes are increasingly higher than the minimum $Q$ factor $ \left( x^2 \, Q \right)_\text{min}$.Both for very small and large gaps the dipole moment in Eq.~\ref{eq:Dipole2cyl} vanishes, in agreement with the diverging $Q$ factor for both small and large gaps in this figure.

So far we have {\it independently} studied the radiation and the dissipation $Q$ factors. In any realistic object these two mechanisms coexist, and the total $Q$ factor is given by
\begin{equation}
    \frac{1}{Q_k} = \frac{1}{Q_k^r} + \frac{1}{Q_k^d}.
\end{equation}
We will now identify scenarios where one of the them is dominant. By taking the limit of Eq.~\ref{eq:Qfac_Cylinders} for $\delta/R\rightarrow 0$, we find that the radiation $Q$ factor of any given mode $k$ diverges as $\displaystyle Q_k^r \approx \frac{4}{\pi}\left(\frac{\delta}{R}\right)^{-\frac{1}{2}} \frac{1}{x^2}$ (as can be seen in Fig.~\ref{fig:Qk_Qmin}). Therefore, for any given value of $\nu$, by decreasing the (relative) gap size $\delta/R$, the material loss will eventually become dominant. Moreover, for a given relative gap size $\delta/R$ there always exists a mode index beyond which the modes are dominated by dissipation loss. In both of those scenarios, the total $Q$ factor of the singular plasmonic structure will eventually be dominated by dissipation.

\begin{figure}[htpb]
\centering
\includegraphics[width=\columnwidth]{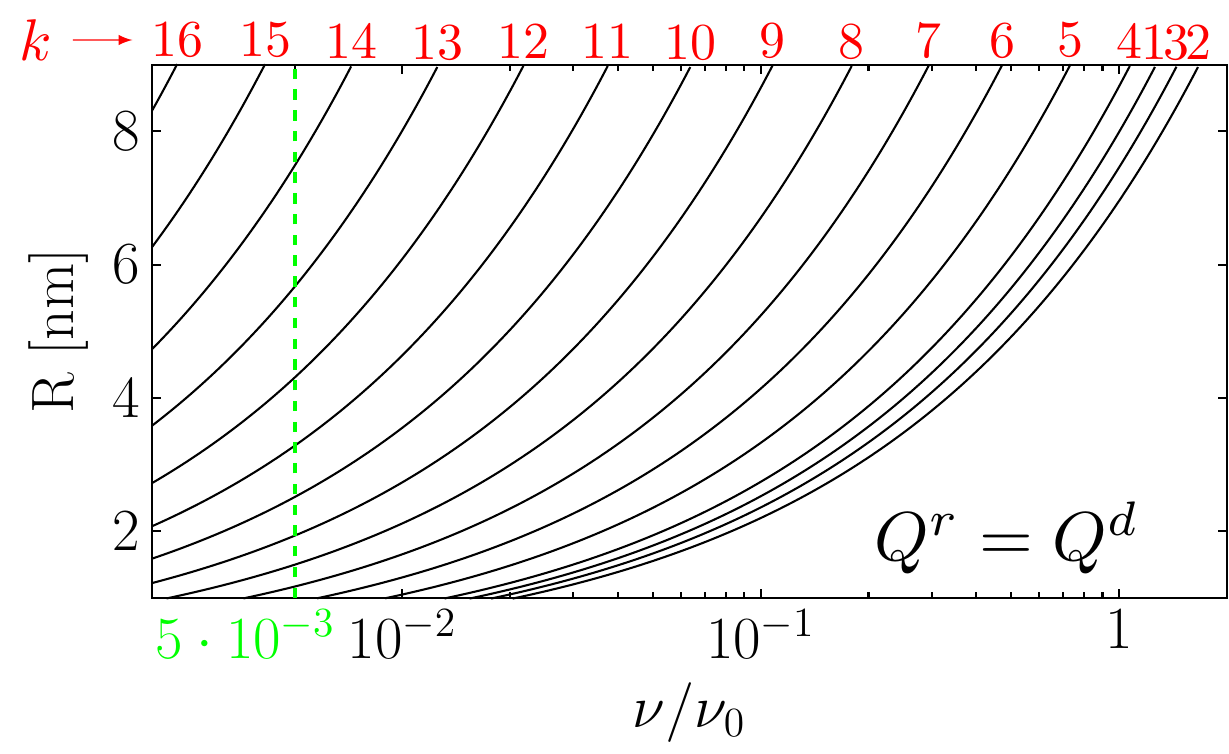}
\caption{Limit curves $\mathcal{R}_k(\nu)$ at $Q^r=Q^d$ in the parameter space $(R,\nu)$ for two Drude metal cylinders at $\delta/R=0.1$, with $\omega_p=13.07\times10^{15}\,{\rm rad/s}$, $\nu_0=131.19\times 10^{12}\, {\rm rad/s}$,  parametrized by the resonant mode index $k$. The curve $\nu/\nu_0 = 5\times10^{-3}$ is also shown (green dashed line).}
\label{fig:QparamC}
\end{figure}

In the most general case, in order to identify the operating regime of the singular plasmonic resonator, it is useful to define for any mode $k$ in the parameter space $(R,\nu)$ the curve in correspondence of which the radiation and the dissipation $Q$ factor are equal $Q_k^d(\nu)=Q_k^r(R)$:
\begin{equation}
\mathcal{R}_k(\nu) = \sqrt{ \frac{2\,c_0^2}{\pi \delta/R\left(4+\delta/R\right)}\frac{{\rm e}^{2 k \mu}-1}{ k }  \frac{\nu}{ {\omega^+_k}^3}
}.
\label{eq:LimitCurve}
\end{equation}
This condition is known as critical coupling and, as we confirm in the Methods, 
when the $k^\text{th}$ resonance is critically coupled $Q_k^d(\nu)=Q_k^r(R)$, its absorption cross section is maximized \cite{haus1984,hamam_coupled-mode_2007,mann_extreme_2013}. Thus, the limit curve $\mathcal{R}_k(\nu)$ coincides with the combination of structure dimension-material loss $(R,\nu)$ at which the singular plasmonic resonator can harvest the greatest amount of power through the $k^\text{th}$ resonance channel. 
The curve $\mathcal{R}_k(\nu)$ also partitions the parameter space $(R,\nu)$ into two regions: one where radiative damping prevails, and one where dissipative damping is dominant. In Fig. \ref{fig:QparamC} we show the curve $\mathcal{R}_k(\nu)$ in the parameter space $(R,\nu)$ for two cylinders at $\delta/R=0.1$, parameterized with the mode index $k$. Depending on the structure dimension and the material loss,  the $Q$-factor may thus fall into three different regimes: 
i) $Q\simeq Q^d\ll Q^r$, i.e., the structure response is dominated by the material loss, as in Figs. \ref{fig:SigmaAbsVaryingDe}(a-c);
ii) $Q\simeq Q^r\ll Q^d$, i.e., the structure response is dominated by radiation loss as in Figs. \ref{fig:SigmaAbsVaryingDe}(d-e);
iii) $
1/Q=1/Q^r+1/Q^d,\,Q^r \sim Q^d$ , i.e., both the material and radiation losses are significant.

As the mode order $k$ increases, the curve $\mathcal{R}_k(\nu)$ shifts toward the upper left corner of the $\left( R, \nu \right)$ plane, and a larger portion of the parameter space is dominated by material loss. The diagram can be more easily understood by fixing the Drude relaxation rate $\nu$: for instance, for $\nu=5\cdot 10^{-3}\nu_0$  (shown with a vertical green line) we deduce that for a  cylinder dimer with $R=5\,$nm, only the first ten modes are dominated by radiation loss, and all other modes are overdamped.


\section{Emergence of a resonance continuum}
\refstepcounter{section}
\label{sec:Link}
We will now turn our attention to the relationship between the broad bandwidth supported by singular plasmonic structures like touching dimers, and the individual modes underlying its spectral response. By reducing the relative gap size, the resonance spectrum of a dimer of nearly touching cylinders transitions from discrete resonances to a continuum. Thus, the modes can no longer be considered isolated, which is why such structures can support a broad resonance bandwidth that appears to violate the Chu limit so dramatically.  
For a given gap size, a broad bandwidth can also be achieved by increasing the material loss: each resonance $Q$ factor decreases, and the corresponding peaks start to overlap. As a drawback, increasing the material loss reduces the structure field enhancement.


\iffig
\begin{figure*}[tb]
\centering
\includegraphics[width=\textwidth]{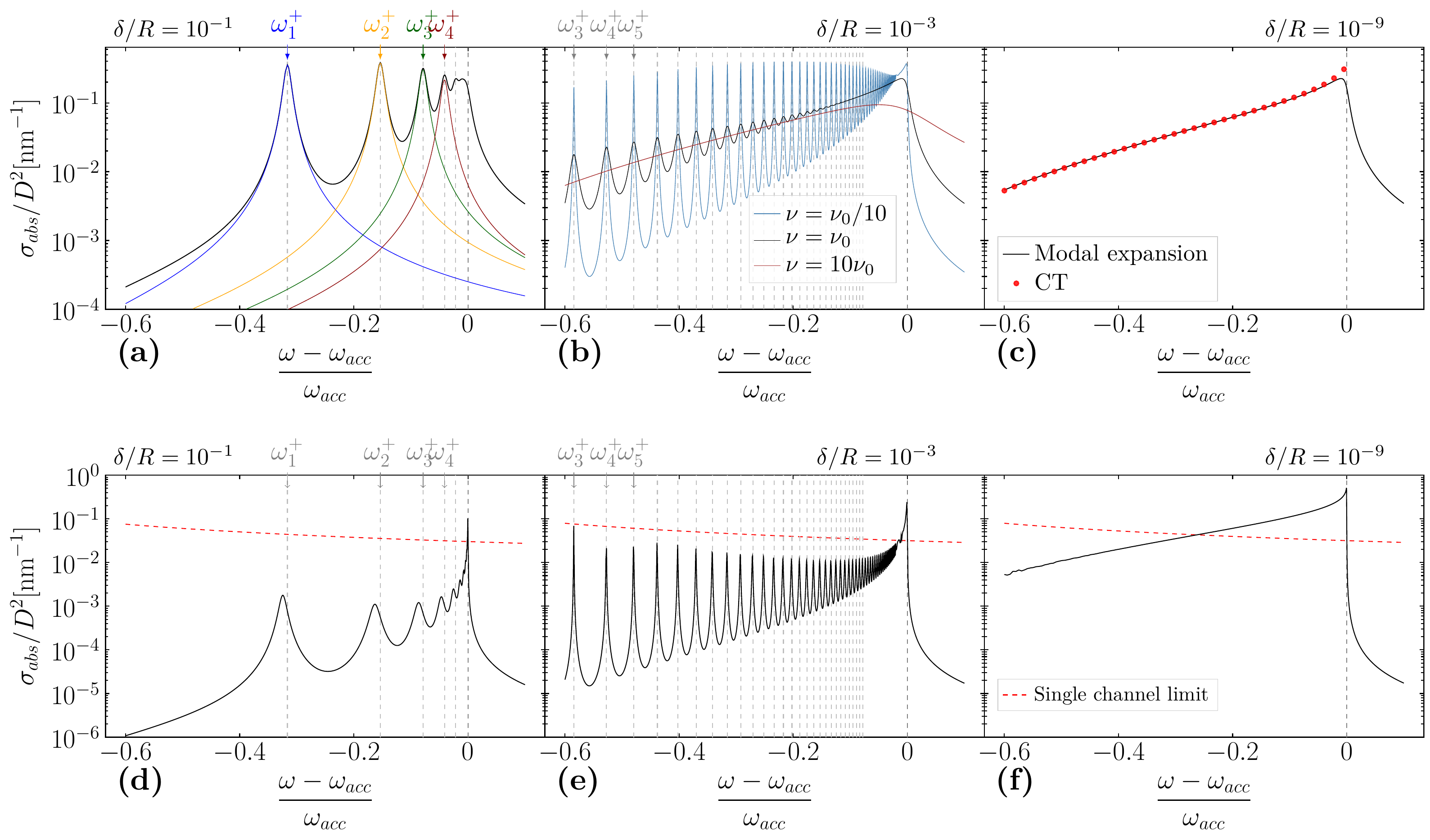}
\caption{Absorption cross-section (normalized by $D^2$) of  a  silver  cylinder dimer excited by a plane wave  polarized along the structure axis, as a function of frequency, in the quasi-electrostatic regime (top panels $\bf a - c$), and for $R=8\,\text{nm}$ (bottom panels $\bf d-f$). $\bf (a)$  Case $\delta/R =10^{-1}$. The partial  absorption cross sections (in color) of the four modes responsible for  the first four peaks are shown. $\bf (b)$ Case  $\delta/R =10^{-3}$. The normalized cross section for the Drude relaxation rate $\nu=\nu_0/10$, $\nu=\nu_0$, and $\nu=10\nu_0$, with $\nu_0=131.19\times 10^{12}\, {\rm rad/s}$,  are shown. In (a-b), the positions of the resonance frequencies are marked by vertical dashed lines. $\bf (c)$ The normalized cross-section for $\delta/R =10^{-9}$ (black line),  calculated using Eq. \ref{eq2},  is compared to the corresponding quantity obtained by conformal transformation approach, valid for $\omega<\omega_{acc}$ \cite{aubry_plasmonic_2010}. The bottom panels correspond to a cylinder dimer with $R=8\,\text{nm}$, $\nu=5\times 10^{-3}\nu_0$, and gaps $\delta/R =10^{-1}$ $\bf (d)$, $\delta/R =10^{-3}$ $\bf (e)$, $\delta/R =10^{-9}$ $\bf (f)$.
 The position of the EQS resonance frequencies and the accumulation frequency are also shown with vertical lines. The dashed red line shows the single channel absorption limit.}
\label{fig:SigmaAbsVaryingDe}
\end{figure*}
\fi

\subsection{Quasi-electrostatic limit}

We first consider a purely electrostatic scenario, where no radiation is considered, and dissipation is the only damping mechanism. We investigate the absorbed power spectrum of two nearly touching cylinders when the gap size is a tenth of the cylinder radius, i.e., $\delta/R=10^{-1}$, assuming a Drude relaxation rate $\nu=\nu_0$. In Fig. \ref{fig:SigmaAbsVaryingDe}(a), we show the absorption cross section with a continuous black line, and the {\it partial} absorption cross section of the first four resonant plasmonic modes using different colors. The {\it partial} absorption cross section is defined as the cross section that we would measure if only one mode were excited at a time, which is calculated using Eq. \ref{eq2} by only considering the $k^\text{th}$ term in the summation. Since in this quasi-electrostatic regime the plasmonic modes are orthogonal, the total $\sigma_\text{abs}$ can be rigorously decomposed in the sum of all the partial ones. Each peak results due to a single mode, thus its fractional bandwidth is the inverse of its dissipation $Q$ factor, whose expression is given in Eq. \ref{eq:QdissDrude}. Near the accumulation point $\omega_{acc}$ the contribution of individual modes can no longer be identified, and the broadening of the curve is due to many closely spaced modes forming a continuum.

We now reduce the gap size to a thousandth of the cylinder radius, i.e., $\delta/R=10^{-3}$, for the same scattering rate $\nu=\nu_0$. We show the corresponding absorption in Fig. \ref{fig:SigmaAbsVaryingDe} (b) (black line). The absorption peaks now spread over the frequency axis since, as $\delta/R$ decreases, the resonance frequencies $\omega_k^+$ undergo a redshift \cite{mcphedran_electrostatic_1981,aubry_plasmonic_2010} (see also Fig.~\ref{fig2}). The same figure also highlights the role of material loss:  $\nu_0/10$ and  $10\nu_0$ are shown in blue and red, respectively. As the material loss increases, each peak broadens consistently with Eq. \ref{eq:QdissDrude}. For high loss $10\nu_0$, the overall absorption curve is very smooth, due to the spectral overlap of adjacent modes. In this case, the individual contribution of plasmon resonances can no longer be identified. In the case of reduced loss, individual modes are discernable much closer to the accumulation point, in agreement with Fig.~\ref{fig:QparamC}. As the material loss increases, the bandwidth of the scattering response is increased, but the field enhancement is reduced, as shown in the Methods.
From this analysis and the considerations made in the previous section, it is apparent that for any given gap size $\delta/R$ there exists a scattering rate value $\nu$, and hence a material loss level, beyond which the peaks start to merge.

Finally, in Fig. \ref{fig:SigmaAbsVaryingDe}(c), we further decrease the gap size to $\delta/R=10^{-9}$, and compare the  absorption cross-section obtained by Eq. \ref{eq2} (black line) with the analytical formula provided in \cite{aubry_plasmonic_2010} (red dots), obtained from a conformal transformation in the limiting case of touching cylinders, and valid for $\omega<\omega_{acc}$. The Drude relaxation rate is $\nu=\nu_0$. The two results are in excellent agreement below the accumulation point and show a smooth absorption spectrum.


\subsection{Beyond the quasi-electrostatic limit}

The analysis carried out in the previous figure, consistent with Ref. \cite{aubry_plasmonic_2010}, was performed in the quasi-electrostatic limit, neglecting radiation. In this regime, however, there is no minimum $Q$ factor and correspondingly a Chu limit, since radiation loss is not present. In order to investigate the role of radiative damping, we now repeat the analysis for a deeply subwavelength structure in the presence of radiation, assuming a very low damping rate $\nu=5\times 10^{-3}\nu_0$.
We adopt the modified long wavelength approximation (MLWA) \cite{meier_enhanced_1983,zeman_accurate_1987,forestiere_resonance_2020}, arresting the expansion of the $k^\text{th}$ bright eigenpermittivity, in the size parameter $x$, around the electrostatic resonance value $\varepsilon_k^+$, as in  Eq. \ref{eq:PertEig} of the Methods.

In Fig. \ref{fig:SigmaAbsVaryingDe}(d) the radius of each cylinder is $R=8\,$nm, and the gap size is fixed at a tenth of the cylinder radius, i.e., $\delta/R=10^{-1}$. We have seen in Fig. \ref{fig:QparamC} that, independent of the value of $\nu$, the $Q$ factor of individual modes is dominated by dissipation for sufficiently high mode numbers. In the present case, the first $13$ peaks are dominated by radiation loss, and the remaining ones by material loss. The radiation $Q$ factor of each of them is greater than the minimum $Q$ achievable by longitudinal currents supported by a pair of cylinders with $\delta/R=10^{-1}$, i.e., $x^2 Q_k \ge x^2 Q_\text{min} = 48/\pi^3 \ge 4/\pi=x^2Q_\text{Chu},\,\forall k$. The red dashed line in this figure shows the single channel absorption limit, which a single mode can only reach when it is critically coupled. Given that all modes are underdamped, the response only exceeds the single channel limit near the accumulation point, where the density of resonances is very high.

By reducing the gap size to $\delta/R=10^{-3}$ (Fig. \ref{fig:SigmaAbsVaryingDe}(e)), the absorption is characterized by denser peaks, spread over the frequency axis. The first mode is seen to approximately reach the single channel limit, and is thus close to critically coupled. All higher order modes are therefore overdamped, in agreement with Eq. \ref{eq:LimitCurve}. As the separation between the cylinders decreases further (Fig. \ref{fig:SigmaAbsVaryingDe}(f)), the resonance frequencies tend to further spread over the frequency axis and eventually form a continuum, as in the pure electrostatic case. The single channel limit is again exceeded because this response cannot be considered due to a single mode anymore.

\begin{figure}[htpb]
\centering
\includegraphics[width=\columnwidth]{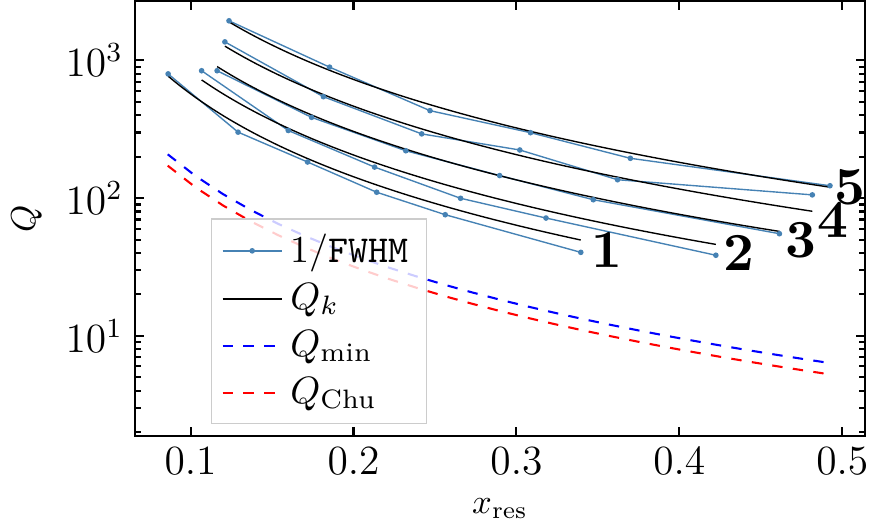}
\caption{Radiation $Q$ factor of the resonant modes responsible for the first five peaks in the absorption cross-section for Drude metal cylinder dimer, with $\nu=5\times10^{-3}\nu_0$, and $\delta/R=0.1$, as a function of the resonant size parameter $x_\text{res}=\frac{\omega_\text{res}}{c_0} R$, calculated as the inverse of the relative full width at half maximum bandwidth by using COMSOL (blue dots) and Eq. \ref{eq:Qfac_Cylinders} (black line). The minimum $Q$ factor achievable by the structure (blue dashed line) and the Chu limit (red dashed line) are also shown.}
\label{fig:Qagreement}
\end{figure}

With the aid of Fig. \ref{fig:Qagreement}, we investigate in greater detail the link between the inverse of the full-width at half maximum ($\texttt{FWHM}^{-1}$) of the first five peaks of the absorption cross section in Fig.~\ref{fig:SigmaAbsVaryingDe}(d), and the radiation $Q$ factor of the first $5$ modes given by Eq. \ref{eq:Qfac_Cylinders}. This analysis is conducted as a function of the size parameter of the structure, considering a  cylinder dimer with $\delta/R=0.1$, and  $\nu = 5\times 10^{-3}\nu_0$, excited by an electric field polarized along the dimer axis. The radiation $Q$ factor calculated using Eq.~\ref{eq:Qfac_Cylinders} shows very good agreement with $\texttt{FWHM}^{-1}$ of the first five peaks. We also plot with a dashed line the Chu limit of a translational invariant radiator $Q_\text{Chu}=\frac{4}{\pi} \frac{1}{x^2}$ (charge density on the minimum circle enclosing the object), and the minimum $Q$ factor for a dimer (blue, dashed). All resonances are characterized by a radiation $Q$ factor greater than the Chu limit, as expected.



\section{Effect of disorder}
\begin{figure}[h!]
\centering
\includegraphics[width=0.95\columnwidth]{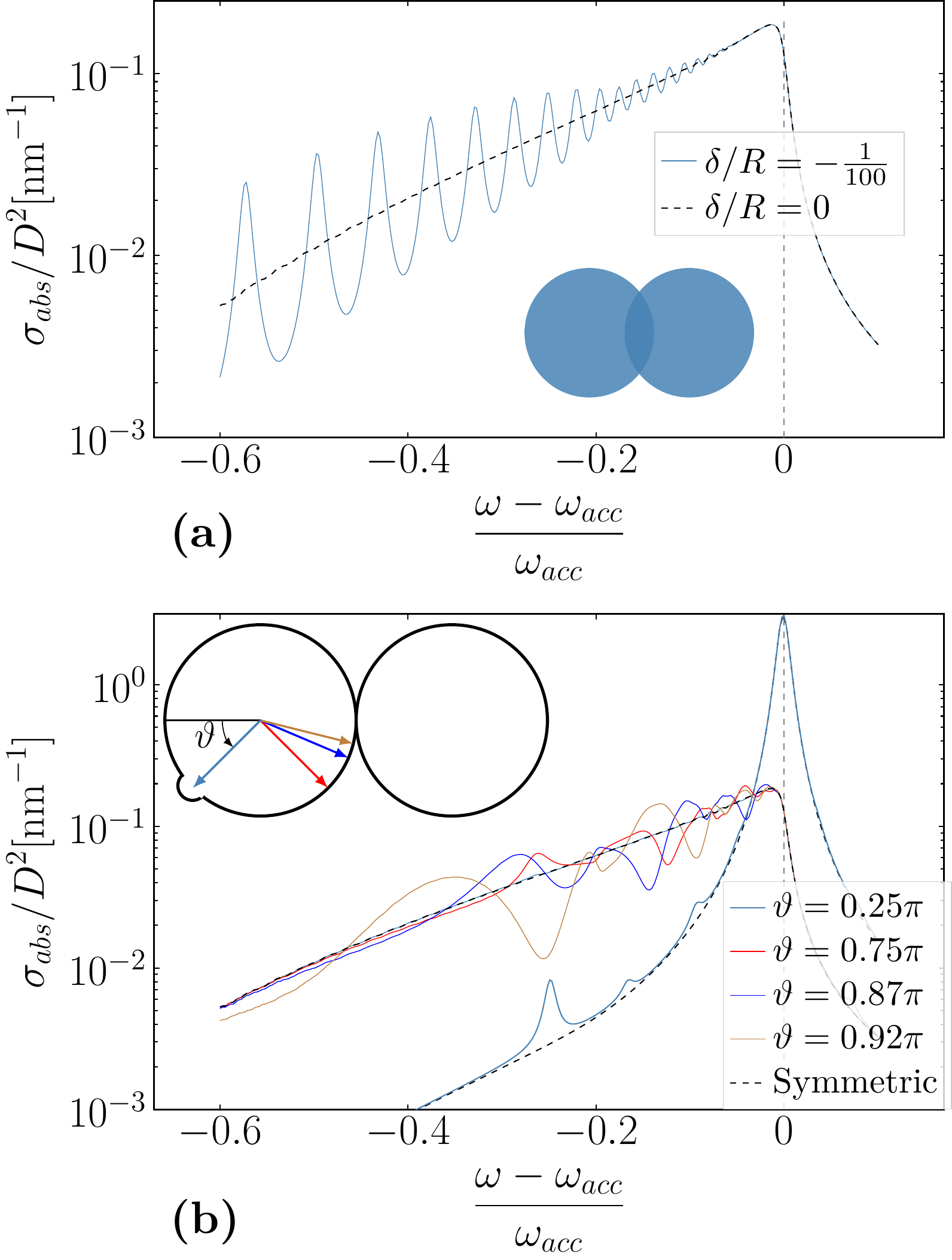}

\caption{{\bf (a)} Absorption cross-section normalized by $D^2$ of a Drude metal cylinder dimer, $\delta/R =-10^{-2}$ (see the inset), compared to the touching cylinders (black dashed line), as a function of frequency. {\bf (b)} Absorption cross-section normalized by $D^2$ of a Drude metal  cylinder dimer with a semicircle of radius $R/25$ added to the cross-section of one of the two cylinders at the angle $\vartheta=0.25\pi,\,0.75\pi,\,0.87\pi,\,0.92\pi$ (see the inset). The parameter $\nu$  is assumed to be $\nu=\nu_0$. The incident field is polarized along the axis of the cylinders. The absorption cross-section normalized by $(2R)^2$ of an isolated symmetric cylinder, and of a cylinder with a semicircle added to its cross-section at the angle $\vartheta=0.25\pi$ (see the inset), excited by an horizontally polarized plane wave, are also shown with a black dashed line and a light-blue solid line, respectively.}
\label{fig8}
\end{figure}

\refstepcounter{section}
\label{sec:Disorder}
Our results so far have demonstrated that the broad bandwidth over which large field enhancements can be obtained in singular plasmonic structures is associated with a large number of densely packed resonances, which individually obey the Chu limit. As a result, despite the fact that the structure is broadband, the stored energy in the system is extremely large at any frequency within the resonance range. Hence, despite their broad bandwidth, these singular structures are expected to be very sensitive to disorder \cite{Johnson_Coupling_2003}, given that a small perturbation on e.g., the surface of the cylinders is exposed to very large field amplitudes. Here, we demonstrate this trade-off based on two possible types of non-idealities: i) two overlapping cylinders and ii) asymmetric cross sections due to surface roughness. In Fig. \ref{fig8}(a) we compare the normalized absorption cross section of two overlapping cylinders intersecting along the dimer axis (as shown in the inset) with $\delta/R = - 10^{-2}$, against the corresponding quantity observed in the ideal (reference) scenario of two touching cylinders.  
Interestingly, the absorption cross section is similar to what we observed in Fig. \ref{fig:SigmaAbsVaryingDe} (b) for $\delta/R = + 10^{-3}$, in other words for overlapping cylinders we are again in the discrete mode regime. The touching configuration thus appears to be an unstable point with respect to variations of the gap size $\delta/R$: a small deviation from the touching case implies a transition from the continuum to the discrete mode regime.


Fig.~\ref{fig8}(b) illustrates the effect of an asymmetry in the cross section of the touching cylinders. Specifically, we include a small \lq \lq bump\rq\rq on the surface, located at the angle $\vartheta$ (see the inset). We compare the corresponding normalized $\sigma_\text{abs}$, parameterized for several values of $\vartheta$ (colored lines), against the ideal case (dashed black line). We conclude that if the bump is placed far from the singularity, i.e., $\vartheta=\pi/4$, there is almost no appreciable effect on the $\sigma_\text{abs}$ spectrum. However, as the irregularity approaches the gap region, the $\sigma_\text{abs}$ curve undergoes strong oscillations. Given that the electric field is strongest near the apex, the stronger effect as $\vartheta$ increases towards $\pi$ is expected. For a single cylinder, supporting only the dipole mode, the effect of a bump on the surface is much less considerable, as can also be observed in Fig.~\ref{fig8}(b).

\section{Conclusions}
\refstepcounter{section}
Singular plasmonic structures exhibit a broad scattering spectrum that at first sight may appear to elude the Chu limit. Here, we have elucidated in detail the relationship between the Chu limit, the modes supported by a singular plasmonic structure, and their radiation and dissipation $Q$ factors. We derived bounds for the radiation $Q$ factors of 2D objects of arbitrary cross-section, and found analytical formulas for the dissipation and radiation $Q$ factor of the bright modes of two nearly touching cylinders. As expected, the radiation $Q$ factor of these modes always exceeds the Chu limit.

Our investigation of the absorption power spectrum of two nearly touching cylinders excited by a plane wave has shed light on the link between the bandwidth of the absorption cross section and the $Q$ factor of each individual mode composing the broad resonance. As long as the peaks of the absorption power spectrum are well separated, their bandwidth is equal to the inverse of the $Q$ factor. If the radiation loss is dominant, the bandwidth of these peaks is also subjected to the Chu limit. However, by reducing the gap size, the resonance spectrum of a cylinder dimer transitions from discrete to continuum. The plasmonic modes are no longer isolated and, in any given spectral interval, the overall absorption response arises from the contribution of multiple modes. Thus, the connection between the $Q$ factor and the bandwidth is lost, and the bandwidth of the absorption power spectrum is no longer subjected to the Chu limit. While a high density of resonances may overcome the trade-off between bandwidth and size of a small radiator, we have shown that this approach comes at the cost of high sensitivity to disorder, since the stored energy in the system remains very large, consistent with the Chu limit.

\section{Methods}
\refstepcounter{section}
\label{sec:Methods}
\label{sec:FieldEnhancement}
\begin{figure*}[ht]
\centering
\includegraphics[width=0.8 \textwidth]{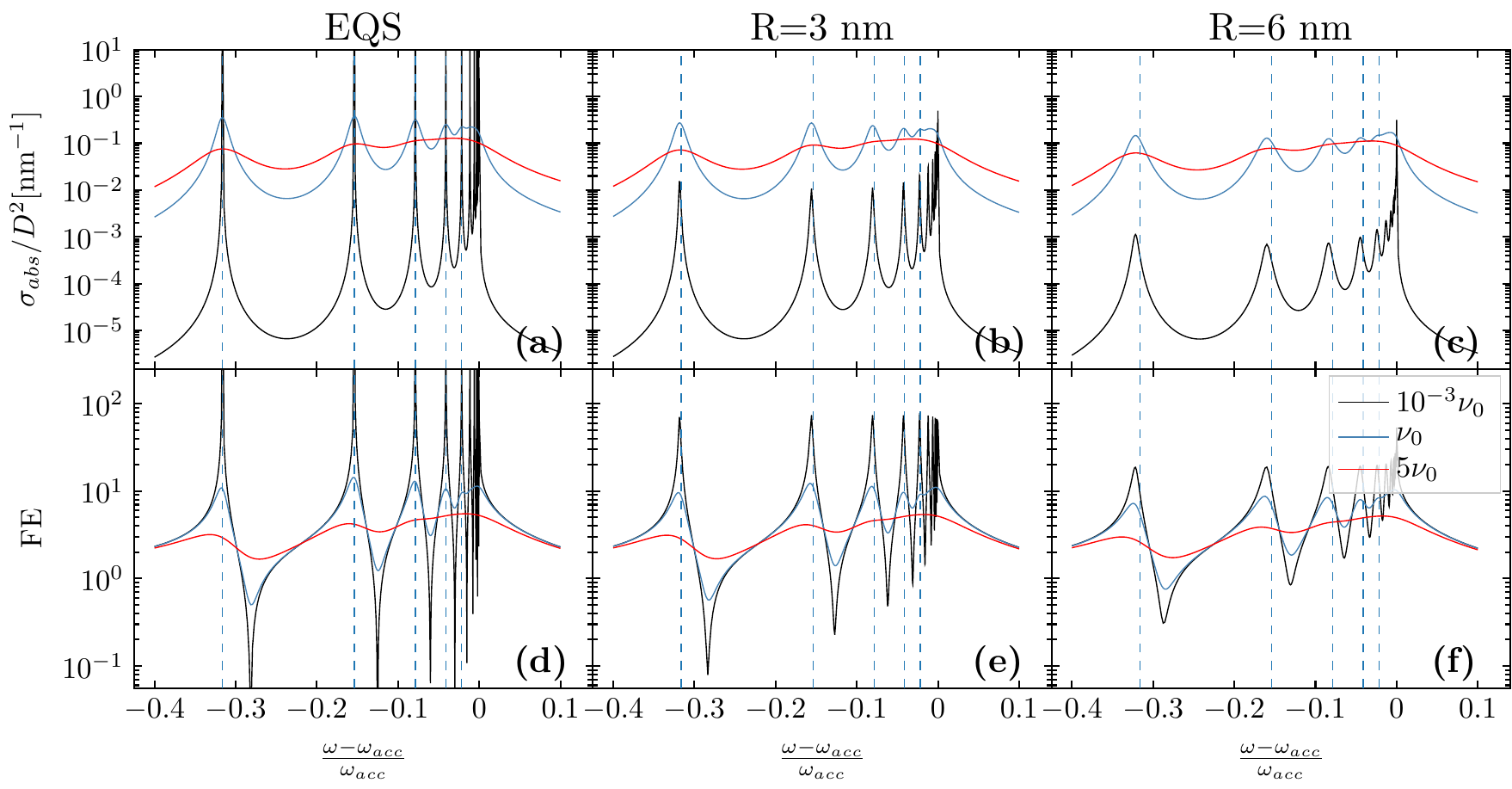}
\caption{Absorption cross-section normalized by $D^2$ $\bf (a-c)$ and field-enhancement $\bf (d-f)$ of  a Drude metal cylinder dimer excited by a plane wave  polarized along the structure's axis, calculated by using the modified long wavelength approximation, as a function of frequency, for $\delta/R=0.1$, {\bf (a)-(d)} $R/\lambda \rightarrow 0$ (quasi-electrostatic regime), {\bf (b)-(e)} $R=3\,$nm, {\bf (c)-(f)} $R=6\,$nm and $\nu=10^{-3} \,\nu_0,\,\,\nu_0,\,\,5 \,\nu_0$, where $\nu_0=131.19\, {\rm Trad/s}$ and $\omega_p=13.07\,{\rm Prad/s}$.}
\label{fig:sigma_losses_Rad}
\end{figure*}

\subsection{Minimum $Q$ for radiators of the electric kind}
\label{sec:Etens}
Following \cite{gustafsson_physical_2015}, here we briefly derive the minimum $Q$ factor for a translational invariant radiator of the electric kind. 

From Eq. \ref{eq:QminEQS}, the minimum $Q$ factor normalized with the squared size parameter $x=\frac{\omega}{c_0} \ell_c$ is defined as the reciprocal of the maximum eigenvalue of the object polarizability tensor ${\boldsymbol \gamma}_e$ scaled by $\ell_c^2$, multiplied by 8. Here $\ell_c$ is a characteristic linear length of the object cross-section (e.g., the radius of the smallest circle enclosing the cross-section). 

The polarizability tensor of a translational invariant geometry is a linear correspondence between an homogeneous external electric displacement field $\varepsilon_0 E_0 \hat{\mathbf{e}}$ and the electric dipole moment $\mathbf{P}$, defined as
\begin{equation}
    \mathbf{P} = \oint_{\partial \Omega} \sigma \rrp \rb \, dl,
\end{equation}
of the charge density distribution $\sigma$  (subjected to the charge neutrality condition $\displaystyle \oint_{\partial \Omega}\sigma dl=0$), solving the linear integral equation \cite{van_bladel_electromagnetic_2007}
\begin{equation}
  -\oint_{\partial \Omega} \sigma \rrpp \frac{ \log \Delta r}{2 \pi} dl' = \left( \varepsilon_0 E_{0} \hat{\boldsymbol{e}} \right)  \cdot \rb \qquad  \forall \rb \in \partial \Omega,
  \label{eq:PolarizabilityE2D}
\end{equation}
where $\Delta r = \left|\rb - \rb '\right|/\ell_c$. Thus, the electric polarizaility tensor ${\boldsymbol \gamma}_e$ is a $2\times 2$ matrix, defined as the map
\begin{equation}
    {\boldsymbol \gamma}_e \cdot \hat{\mathbf{e}} \varepsilon_0 E_0 = \mathbf{P}.
\end{equation}

When $\Omega$ is a circular cylinder section of radius $R$, and characteristic linear length $\ell_c=R$, we have $\sigma = 2\varepsilon_0 E_0 \hat{\mathbf{e}}\cdot \hat{\rb},\,\,\forall \rb \in \partial\Omega$, with $\mathbf{P} = 2\pi R^2 \varepsilon_0 E_0 \hat{\mathbf{e}}$. Thus, $\boldsymbol{\gamma}_e = 2\pi R^2 I$, being $I$ the identity matrix, and the minimum $Q$ factor is given in Eq. \ref{eq:minQEQS}.

\subsection{Minimum $Q$ for radiators of magnetic kind}\label{sec:Mtens}
Following \cite{gustafsson_physical_2015}, here we briefly derive the minimum $Q$ factor for a translational invariant radiator of the magnetic kind.

The magnetic polarizability tensor $\boldsymbol{\gamma}_m$ of a translational invariant object, with characteristic linear length $\ell_c$, is a linear correspondence between an homogeneous external magnetic field $H_0 \hat{\mathbf{h}}$  and the magnetic dipole moment, defined as:
\begin{equation}
    \mathbf{M} = \frac{1}{2} \int_{\Omega} \rb \times \mathbf{j} \,dS,
\end{equation}
where $\mathbf{j}$ is a current density distribution having zero-average over $\Omega$ and solving the integral equation problem \cite{van_bladel_electromagnetic_2007}:
\begin{equation}
 -\int_{\tOmega} \mathbf{j} \rrpp \frac{ \log \Delta r}{2\pi} dS' = \frac{1}{2} H_{0} \hat{\mathbf{e}} \times \rb, \quad \forall \rb \in \Omega,
\label{eq:PolarizabilityM3D}
\end{equation}
with $\Delta r = \left|\rb - \rb '\right|/\ell_c$. Thus, the magnetic polarizability tensor ${\bf \gamma}_m$ is a scalar, defined as the map
\begin{equation}
    {\bf \gamma}_m  \hat{\bf h} H_0 = {\bf M}.
\end{equation}

When $\Omega$ is a circular cylinder section of radius $R$, and characteristic linear length $\ell_c=R$, we have ${\bf j} = H_0 \delta(r-R)\,\hat{\bf z}\times \hat{\rb}$, where $z$ is the cylinder axis direction, and $\delta(r-R)$ is the Dirac delta-function that is 0 everywhere except for $r=R$, i.e., the current yelding the minimum $Q$ is a current loop localized on the cylinder boundary. Moreover, we have ${\bf M} = \pi R^2 H_0$, and hence $\gamma_m = \pi R^2$. The minimum $Q$ factor is equal to the one for radiators of the electric type, given in Eq. \ref{eq:minQEQS}.
\subsection{Quasi-electrostatic Modes}
\label{appendix:Modes}
The normalized quasi-electrostatic modes of a cylinder dimer with gap size $\delta$ and radius $R$  have the following expression:
\begin{multline}\label{eq:modes_expression}
{\bf E}_k =R\sqrt{\frac{k}{8\pi}} \left(\cosh{u}-\cos{v}\right)\\
\left\{  \begin{aligned}
  \displaystyle  & e^{-k\,u}\left(e^{2k\mu}-1\right)\left(\hat{\bf v}\sin{k v}+\hat{\bf u}\cos{k v}\right), & \\
  & \qquad\qquad\qquad \qquad\qquad\qquad\qquad u\ge\mu &\\
  \displaystyle   & 2 \left(\hat{\bf v}\sin{k v}\,\sinh{k u}-\hat{\bf u}\cos{k v}\,\cosh{k u}\right) , & \\
  & \qquad\qquad\qquad  \qquad\qquad\qquad-\mu\leq u\leq\mu &\\
  \displaystyle  & e^{k\,u}\left(e^{2k\mu}-1\right)\left(-\hat{\bf v}\sin{k v}+\hat{\bf u}\cos{k v}\right), &\\
  &  \qquad\qquad\qquad  \qquad\qquad\qquad\qquad\mu\leq-\mu &
  \end{aligned}\right.
\end{multline}
where  $-\infty\leq u \leq \infty,\,0 \leq v <  2\pi$ are the bipolar coordinates. The circular boundaries of the cylinder dimer coincide with lines $u=\mu$ and $u=-\mu$, where $\mu=\arccosh{\left(1+\frac{\delta}{2R}\right)}$, being $\arccosh$ the inverse hyperbolic cosine \cite{mayergoyz_analysis_2007}.

\subsection{Absorbed Power and Radiation Corrections}\label{appendix:Pabs_RadCorr}
Within the quasi-electrostatic approximation, the absorbed power  per unit length $P_{abs}$ of a cylinder dimer illuminated by a plane wave of amplitude $E_0$ and linearly polarized along the system axis $\hat{\bf x}$ is 
\begin{multline}
\label{eq:PabsMIM}
    P_{abs}  =\frac{1}{2} \varepsilon_0   \text{Im} \left\{ \varepsilon_R \right\}\omega  \int_\Omega  \left| {\bf E}_{tot} \right|^2 d S= \\ \frac{1}{2} \varepsilon_0   \text{Im} \left\{ \varepsilon_R  \right\}\omega  \left|E_0\right|^2\sum_{k=1}^{\infty} \left|\frac{\varepsilon_k-1}{\varepsilon_k-\varepsilon_R}\right|^2 \left\langle \hat{\bf x},{\bf E}_k\right\rangle^2.
\end{multline}

In order to include the radiation loss in this framework, assuming the object electrically small, i.e., $x<1$, we adopt the modified long wavelength approximation (MLWA) \cite{meier_enhanced_1983,zeman_accurate_1987,forestiere_material-independent_2016} arresting the expansion of the $k^\text{th}$ eigenpermittivity $\epsilon_k(x)$   around the electrostatic resonance to the first real and imaginary corrections, yielding
\begin{multline}
\label{eq:PertEig}
\epsilon_k\left(x\right)\simeq \varepsilon_k - \frac{\left(\varepsilon_k-1\right)^2}{4\pi}\left|{\bf P}_{{k}}\right|^2 \left(- \log{x} +{\rm i}\,\frac{\pi}{2}\right)x^2\\
=-\coth {k\mu}  -\frac{2\,\delta/R}{\delta/R+4} \frac{k}{\sinh^2{k\mu}} \left(- \log{x} + {\rm i}\,\frac{\pi}{2}\right)x^2.
\end{multline}
\subsection{Critical Coupling}\label{sec:crit_coup}
Within the framework of the quasi-electrostatic mode expansion, we show that the absorption cross-section of any translational invariant object reaches a maximum on resonance, when the dissipation $Q$ factor is equal to the  radiation $Q$ factor ($Q^d = Q^r$), in agreement with \cite{hamam_coupled-mode_2007}, in which the coupled-mode theory formalism is employed.

By using Eq. \ref{eq:Qfac_Cylinders}, the expansion of the eigenpermittivity $\epsilon_k $ in Eq. \ref{eq:PertEig} at the resonance of the $k^\text{th}$ mode (for the coupled cylinders, $x = \frac{\omega_k}{c_0}\frac{D}{2} $), can be rewritten as
\begin{equation}\label{eq:eig_exp_Q}
    \epsilon_k (\omega_k)\simeq \varepsilon_k +{\rm i} \frac{\varepsilon_k-1}{Q_k^r}.
\end{equation}
The relative dielectric perimittivity $\varepsilon_R$ in \eqref{eq:Drude} at the resonance of the $k^\text{th}$ mode, for which $\mbox{Re}\{\varepsilon_R(\omega_k)\} = \varepsilon_k$, can be recast as
\begin{equation}\label{eq:drude_res}
    \varepsilon_R(\omega_k) = \varepsilon_k - {\rm i} \frac{\varepsilon_k-1}{Q_k^d},
\end{equation}
where $Q_k^d$ is given in Eq. \ref{eq:QdissDrude}. By plugging Eqs. \ref{eq:eig_exp_Q}- \ref{eq:drude_res} in  Eq. \ref{eq:PabsMIM}, and normalizing by the incoming irradiance ($ c_0 \varepsilon_0 E_0^2/2 $), the $k^\text{th}$ partial absorption cross-section $\sigma_{\text{abs}_k}$(on $k^\text{th}$ resonance) can be expressed as
\begin{multline}
   \sigma_{\text{abs}_k}= \left\langle \hat{\bf x},{\bf E}_k\right\rangle^2  \left.\frac{\omega \mbox{Im}\{\varepsilon_R\}|\epsilon_k-1|^2}{ \mbox{Im}\{\epsilon_k  - \varepsilon_R \}^2}\right|_{\omega=\omega_k}\\
    =\left\langle \hat{\bf x},{\bf E}_k\right\rangle^2 \omega_k (1-\varepsilon_k)\frac{Q_k^d \left[1+\left(Q_k^r\right)^2\right]}{\left(Q_k^d+Q_k^r\right)^2}.
\end{multline}
The condition $Q_k^d=Q_k^r$ is a saddle point for the two-variable function $\displaystyle g(Q_k^d,Q_k^r)=\frac{Q_k^d \left[1+\left(Q_k^r\right)^2\right]}{\left(Q_k^d+Q_k^r\right)^2}$: by fixing the structure dimension, and therefore $Q_k^r$, the function reaches its maximum for $Q_k^d=Q_k^r$; by fixing the material loss, and therefore $Q_k^d$, the function grows with $Q_k^r$, and hence decreases with the structure dimension.
\subsection{Field Enhancement}
In Sec. \ref{sec:Link}, we investigated the effect of an increase of material loss to the absorbed power spectrum of the cylinder dimer. In particular, it was shown that by increasing the level of loss, the overall absorption curve becomes very smooth, due to the spectral overlap of adjacent modes. Here, we investigate the corresponding consequences on the field enhancement. The field enhancement is defined as the ratio between the average electric field inside the two cylinders and the plane wave amplitude.

In the quasi-electrostatic regime, as the material loss tends to zero, the absorbed power spectrum, shown in Fig. \ref{fig:sigma_losses_Rad}  (a), becomes a series of impulse functions centered at the resonance frequencies. Conversely, increasing the material loss, the absorption power spectrum becomes broader. However, as documented in Fig. \ref{fig:sigma_losses_Rad}  (d),  higher material loss leads to a lower field enhancement, which may undermine  - depending on the application of choice - the capabilities of the resonator. This claim remains true also when finite size structures are considered and the radiation included, as shown in Fig. \ref{fig:sigma_losses_Rad} (b),(e) for  $R=3\,$nm, and in Fig. \ref{fig:sigma_losses_Rad} (c),(f) for $R=6\,$nm.

\providecommand{\latin}[1]{#1}
\makeatletter
\providecommand{\doi}
  {\begingroup\let\do\@makeother\dospecials
  \catcode`\{=1 \catcode`\}=2 \doi@aux}
\providecommand{\doi@aux}[1]{\endgroup\texttt{#1}}
\makeatother
\providecommand*\mcitethebibliography{\thebibliography}
\csname @ifundefined\endcsname{endmcitethebibliography}
  {\let\endmcitethebibliography\endthebibliography}{}

\end{document}